\documentclass{ws-rv975x65}
\usepackage{ws-rv-van}             
\makeindex
\begin{document}

\chapter{Finite Size Scaling Analysis of the Anderson Transition\label{ch1}}

\author[B. Kramer, A. MacKinnon, T. Ohtsuki and K. Slevin]{B. Kramer${^1}$, A. MacKinnon${^2}$, T. Ohtsuki${^3}$ 
and K. Slevin${^4}$}
\address{${^1}$School of Engineering and Sciences, Jacobs University Bremen,\\
Campus Ring 1, 28759 Bremen, Germany\footnote{corresponding author}\footnote{permanent affiliation: 
1. Institut f\"ur Theoretische Physik, Universit\"at Hamburg, Jungiusstra\ss{}e 9, 20355 Hamburg, Germany.}\\
b.kramer@jacobs-university.de}
\address{${^2}$Blackett Laboratory, Imperial College London,\\
South Kensington Campus, London SW7 2AZ, UK.}
\address{${^3}$Physics Department, Sophia University\\
Kioi-cho 7-1, Chiyoda-ku, Tokyo, Japan.}
\address{${^4}$Department of Physics, Graduate School of Science,
Osaka University, 1-1 Machikaneyama, Toyonaka, Osaka 560-0043, Japan.\\
slevin@phys.sci.osaka-u.ac.jp}

\begin{abstract}
  This chapter describes the progress made during the past three
  decades in the finite size scaling analysis of the critical
  phenomena of the Anderson transition. 
  The scaling theory of localisation and the Anderson model of localisation are briefly sketched. 
  The finite size scaling method is described. Recent results
  for the critical exponents of the different symmetry classes
  are summarised.
  The importance of corrections to scaling are emphasised. 
  A comparison with experiment is made, and a direction for future
  work is suggested.
\end{abstract}

\body

\section{Introduction}\label{sec1.1}
Originally, the phenomenon of localisation is a property of
quantum mechanical wave functions bound in potential wells of finite
range. At infinity, where the potential vanishes, the wave functions decay
exponentially for negative energies indicating that the probability of
finding the particle far from the potential well vanishes. This is
called ``potential localisation". It had already been suggested in the 1950s
that potentials with infinite range can also support the existence of 
localised wave functions at positive energies provided that the spatial variation of the potential
is random. This localisation phenomenon is due to destructive
interference of randomly scattered partial waves and is now referred to as
``Anderson localisation''. The most important physical
consequence of Anderson localisation is the suppression of
diffusion at zero temperature, which was conjectured 
by P.W. Anderson in his seminal paper.\cite{pwa1958}
Perhaps, the most striking example of Anderson localisation is in
one dimensional random potentials where all the states are localised, irrespective of
their energy. The study of one dimensional localisation was
pioneered by Mott and Twose \cite{mt1961} and by Gertsenshtein and Vasilev.\cite{gv1959}
It can be treated exactly and has been the subject of several
reviews.\cite{i1973,ar1978,eh1982}

In higher dimensions the problem is more subtle, with the possibility of
energy regions corresponding to localised states only, and to extended
states only, separated by critical energies, called ``mobility
edges''.
The zero temperature and zero frequency electrical
conductivity $\sigma_{0}$ of the system vanishes if the Fermi energy
is located in a region of localised states. In the region of extended
states, $\sigma_{0}\neq 0$. In the absence of interactions, the system
is an electrical insulator in the former case while in the latter case
metallic conductivity is expected.
It was conjectured in a seminal work\cite{aalr1979}
that this metal-insulator transition exists only
in three dimensions, while in dimensions $d\leq 2$ systems are always
insulating. This conjecture was based on the hypothesis of one parameter
scaling of the conductance $g(L)$ of a system of size $L$, i.e. that the dependence
of the conductance on system size can be described by a beta-function,
\begin{equation}
\beta(g)=\frac{{\rm d}\,{\rm ln}\,g(L)}{{\rm d}\,{\rm ln}\,L}\,,
\label{eq1.1}
\end{equation}
that depends only on the conductance. The behaviour of $\beta (g)$ with
$g$ was conjectured based on perturbation theory in the
limits of weak and strong disorder (large and small conductance), and
assuming continuity and monotonicity in between.
Moreover, according to the scaling theory, at the mobility edge a continuous quantum
phase transition between an insulator and a metal occurs accompanied by
the power law behaviour of physical quantities, described by critical exponents, 
that is typical of critical phenomena at continuous phase transitions. 
The critical exponents of the conductivity
\begin{equation}
\sigma_{0}\sim (E-E_{0})^{s}\,,
\label{eq:conductivityexponent}
\end{equation}
and the localisation length
\begin{equation}
\xi \sim (E_{0}-E)^{-\nu}\,,
\end{equation} 
were predicted to obey Wegner's previously conjectured scaling law \cite{wegner79}
\begin{equation}
s=(d-2)\nu\,.
\label{eq:wegner}
\end{equation}
While this work was a great leap forward in our understanding of Anderson localisation,
it remained to establish the validity of the central assumption of the theory,
namely the one parameter scaling hypothesis.

This question was addressed numerically by simulating the Anderson model\cite{pwa1958} of
disordered quantum systems which consists of a
delocalising kinetic energy modelled by a hopping term $V$ and a
localising random potential energy $\epsilon_{j}$, commonly assuming a
white noise distribution of width $W$, on a discrete square lattice
$\{j\}$,
\begin{equation}
H=V\sum_{j,\delta}\mid j\rangle\langle j+\delta\mid +
 \sum_{j}\epsilon_{j}\mid j\rangle\langle j\mid \,,
\label{eq1.2}
\end{equation}
where $\delta$ denotes the nearest neighbours of the lattice site $j$.
Such simulations allowed the one parameter scaling hypothesis to be verified
with a reasonable numerical precision,\cite{mk1981,mk1983,ps81,ps81a}
in the center of the band, at energy $E=0$, and also
to confirm the prediction $s=\nu.$\cite{mk1981}
The critical disorder in three dimensions was
initially found to be $W_{c}(E=0)=16\pm 0.5$ while
$s=\nu=1.2\pm 0.3$.
Although this latter value seemed to be consistent with $\nu=1$
there were subsequently substantial doubts about whether or not this was indeed the
case. It was found necessary to improve the precision of the estimate of the
critical exponent and to study in detail and with high precision the
conditions for the validity of the one parameter scaling hypothesis. Later, it was
found that the exponent was in fact \emph{not} unity and this
intriguing discrepancy was the reason for numerous further numerical
as well as analytical efforts, especially since the experimental
situation was also far from clear.\cite{itoh04}

In the following sections, we briefly review the development of the finite size scaling 
analysis of the Anderson transition, paying particular attention to the role of symmetry and
the estimation of the critical exponents.
We stress the importance of the taking proper account of corrections to scaling,
which has been found to be essential in order to estimate the critical exponents precisely. 
Finally, we tabulate the ``state of the art" estimates for the critical exponents of the 
different universality classes.

Some of the early results have been described in previous review
articles.\cite{km1993,kok2005,h1995,em2008}

\section{The Anderson Model of disordered systems}
\label{sec1.2}
In this section we briefly explain the Anderson model of localisation.
We generalise Eq.~(\ref{eq1.2}) in order to describe more general
systems with different symmetries.
The most general form of Eq.~(\ref{eq1.2}) is 
\begin{equation}
H=\sum_{j\mu, j'\mu'}V_{j\mu, j'\mu'}\mid j\mu\rangle\langle j'\mu'\mid +
\sum_{j\mu}\epsilon_{j\mu}\mid j\mu\rangle\langle j\mu\mid\,.
\label{eq1.3}
\end{equation}
The states $\mid j\mu\rangle$ that are associated with the sites of a
regular lattice $j$ --- usually for simplicity a square lattice is
assumed --- are assumed to form a complete set such that $\langle
j\mu\mid j'\mu'\rangle = \delta_{j,j'}\delta_{\mu,\mu'}$. Indices
$\mu$ denote additional degrees of freedom associated with the lattice
sites which lead to several states per site. If there are $n$ states
the above Hamiltonian describes Wegner's $n$-orbital model.\cite{wegner79,w1979a}
In general, the potential energies $\epsilon_{j\mu}$ and the hopping
integrals $V_{j\mu, j'\mu'}$ are random variables described by some
statistical distributions.

If the energy bands emerging due to the broadening by the kinetic terms 
are not strongly overlapping, we may use the single band approximation 
Eq.~(\ref{eq1.2}). 
In addition, if we assume that sufficiently close to the Anderson critical point
the critical phenomena are \emph{universal},
i.e. independent of the microscopic details of the system, 
Eq.~(\ref{eq1.2}) is the simplest model that can describe the
critical behaviour at the Anderson transition.
In principle, these assumptions
have to be verified \emph{a posteriori}, and to some extent this
has indeed been done during the past decades. 

If the Anderson transition is a genuine phase transition, the critical
behaviour can be expected to depend only on symmetry and dimensionality. 
For a disordered system, spatial symmetry is absent and only
two important symmetries remain: invariance with
respect to time reversal, and invariance with respect to spin
rotations.
Three symmetry classes are distinguished\footnote{In fact,
the classification is more complicated.\cite{altland97,kok2005,em2008}
However, for the present purposes, the following classification is sufficient.}:
the orthogonal class which is invariant with respect
to both time reversal and spin rotations,
the symplectic class which is invariant with respect to time reversal
but where spin rotation symmetry is broken,
and the unitary class where time reversal symmetry is broken.
Note that, if time reversal symmetry is broken, the system is classified
as unitary irrespective of its invariance, or otherwise, 
under spin-rotations.

When the kinetic energy parameter $V$ is a real number, the Anderson model Eq.~(\ref{eq1.2})
is time reversal invariant and belongs to the \emph{orthogonal class}.
In this case, universality has been verified by showing that a Gaussian, Cauchy and
a box distribution of the disorder potential give the same critical
exponents.\cite{so1999}

When the kinetic terms $V_{j, j'}$ become complex, the system
is no longer time reversal invariant and thus belongs to the \emph{unitary
class}. This can be physically realised by applying a magnetic field.
Then, the hopping term has to be replaced by the Peierls substitution
\begin{equation}
V_{jj'}=V\exp{\Large[{\,\rm i}\frac{e}{\hbar}
\int_{j}^{j'}{\bf A}\cdot{\rm d}{\rm x}\Large]} \,,
\label{eq1.4}
\end{equation}
where the vector potential $\bf A$ describes the magnetic field, ${\bf
  B}=\nabla\times {\bf A}$. Two
different unitary models can be constructed using the Peierls Hamiltonian, namely a
\emph{random phase model} which is characterised by
\begin{equation}
V_{jj'}=V\exp{({\rm i}\varphi_{jj'})}\,,
\label{eq1.5}
\end{equation}
with the uncorrelated phases $\varphi_{jj'}$ as random variables, and
a model of a uniform magnetic field that leads to a similar
expression for the kinetic term but with \emph{correlated phases}. Whether or
not these two unitary models have the same critical behaviour has
been the subject of numerous studies.

In the presence of spin-orbit interaction, spin rotation symmetry is broken.
The simplest Hamiltionian for such a \emph{symplectic} case is\cite{aso2002,aso2004,oyo2003}
\begin{equation}
H= \sum_{j}\epsilon_{j}\mid j\rangle\langle j\mid + 
V\sum_{jj'}U_{j,j'}
\mid j\rangle\langle j'\mid \,,
\label{eq1.6}
\end{equation}
where $U_{j,j'}$ is an SU(2) matrix.
This model describes a two dimensional electron system in the presence of
Rashba\cite{br1984} and Dresselhaus\cite{d1955} spin-orbit couplings.
 
If the Anderson transition is a genuine quantum phase transition, we expect that
the critical behaviour is universal and
that the critical exponents depend only on the symmetry class and the dimensionality.

\section{Finite size scaling analysis of the Anderson transition}
\label{sec1.3}

In principle, phase transitions occur only in the thermodynamic limit, i.e. in an infinite system.
In practice, computer simulations are limited to systems of small size.
This necessitates an extrapolation to the thermodynamic limit.
This extrapolation is far from trivial. It requires a numerically stable procedure
which, at least in principle, allows control of the
errors involved. This is especially the case when
the goal is precise estimates of the critical exponents. Finite size scaling is such a procedure.

\subsection{Finite size scaling}

The raw data for the finite size scaling procedure is some appropriate physical quantity in a system of 
finite size.
For some physical quantities it may be necessary to take a statistical average.
An example is the two terminal conductance where
an average over a large number of realisations of the random potential is required.
For self-averaging quantities an average may not be required. 
An example is the quasi-one dimensional localisation length of the electrons on a very long bar where
simulation of a single realisation is sufficient.

This physical quantity $\Gamma$ to be analysed depends on the system size $L$ and a set of parameters $\{w_{i}\}$
\begin{equation}
\Gamma = \Gamma(\{w_{i}\}),L) \,.
\label{eq1.10}
\end{equation}
These latter parameters characterise the distribution function of the potential energies and also
other system parameters such as the energy $E$, applied magnetic field ${\bf B}$, 
spin-orbit couplings, etc.
The extrapolation to the thermodynamic limit is performed by assuming that
that $\Gamma$ obeys a scaling law
\begin{equation}
\Gamma = F(\chi L^{1/\nu}, \phi_{1}L^{y_{1}}, \phi_{2}L^{y_{2}},\ldots)\,.
\label{eq1.12}
\end{equation}
Here, for the sake of simplicity, we assume that $\Gamma$ is dimensionless.
The hope is that, in the thermodynamic limit, only one of
the many scaling variables $(\chi, \phi_{1}, \phi_{2}, \ldots)$ turns
out to be relevant, say $\chi$, and the others $\{\phi_{i}\}$ irrelevant.
Here, the words relevant and irrelevant are used in the technical sense that 
the exponent of the relevant scaling variable is positive $\nu>0$ and the
exponents of the irrelevant scaling variables are negative $y_{i}<0$.
This \emph{ad hoc} assumption has, of course, to be
verified during the numerical analysis.

For very large systems the contribution of the irrelevant scaling variables can be 
neglected and we obtain a one parameter scaling law
\begin{equation}
\Gamma = f(L/\xi)\,,
\label{eq1.13}
\end{equation}
with a correlation length,
\begin{equation}
\xi \sim  |\chi|^{-\nu}\,,
\end{equation}
that depends on the parameters $\{w_{i}\}$. This limit is rarely reached in numerical
simulations and we are forced to deal with the corrections to this one parameter
scaling behaviour due to the irrelevant scaling variables. (Below we shall refer rather
loosely to ``corrections to scaling"; strictly speaking we mean corrections to one parameter scaling.)

In practice, we need to simulate not too small
systems such that consideration of at most
one irrelevant scaling variable is sufficient. In this case, the
scaling form Eq.~(\ref{eq1.12}) reduces to
\begin{equation}
\Gamma = F(\chi L^{1/\nu},\phi L^{y})\,.
\label{eq1.25}
\end{equation} 
We then fit numerical data for the region close to the phase transition by
Taylor expanding the scaling function and the scaling variables, and performing
a non-linear least squares fit.
It is important to control the errors in this fitting procedure carefully and to 
specify the precision of all numerical estimates, if the results are to be scientifically meaningful.
For details we refer the reader to  the article by Slevin and Ohtsuki.\cite{so1999}

Such finite size scaling analyses have been used successfully to analyse the Anderson transition 
in three dimensional systems in various symmetry classes,\cite{so1999,so1997,smo2001,asada05}
the Anderson transition in two dimensional systems with spin-orbit coupling \cite{aso2002,aso2004}
and the plateau transition in the integer quantum Hall effect.\cite{slevin09}

\subsection{Quasi-one dimensional localisation length}

The next question is which physical quantity to use in the finite
size scaling analysis. 
It must be sensitive to the nature, localised or extended,
of the eigenstates. (This rules out the average of the density of states, for example.)
It should also be easily determined numerically with a high precision. 
There are several possibilities. 
One is the localisation length of electrons on a very long bar. 
Another possibility is the level spacing distribution.\cite{shklovskii93,zk1997}
Yet another possibility is the Landauer conductance of a hypercube.\cite{smo2001,smo2003} 
In this section we discuss the first of these possibilities in detail.

Consider a very long  $d$-dimensional bar with linear cross-section $L$. 
This is a quasi-one dimensional system in which all states, irrespective of
the values of the parameters $w_{i}$
are known to be exponentially localised with a quasi-one dimensional localisation
length $\lambda(L;w_{1},w_{2},\ldots)$.
Using this quasi-one dimensional localisation length we define a dimensionless quantity, 
sometimes called the MacKinnon-Kramer parameter,
\begin{equation}
\Lambda(L;w_{1},w_{2},\ldots)=\frac{\lambda(L;w_{1},w_{2},\ldots)}{L}\,.
\label{eq1.15}
\end{equation}
In practice, the error analysis of the simulation is simplified by working directly with the inverse of the 
MacKinnon-Kramer parameter
\begin{equation}
\Gamma = \Lambda^{-1}.
\end{equation}
In the localised phase, $\Gamma$ increases with $L$ for large enough $L$, while in 
the extended phase, it decreases.
Exactly at the critical point we have scale invariance for sufficiently large  $L$
\begin{equation}
\lim_{L\to\infty}\,\Gamma(L)\,={\rm const}=\Gamma_\mathrm{c}\,.
\label{eq1.18}
\end{equation}

\subsection{The transfer matrix method}

The transfer matrix method is the most efficient way of calculating the quasi-one 
dimensional localisation length.\cite{mk1981,km1993}
The Schr\"odinger equation for the Anderson Hamiltonian on a $d$-dimensional bar is rewritten as
\begin{equation}
{\bf V}_{n,n+1}{\bf a}_{n+1}=(E-{\bf H}_{n}){\bf a}_{n}-{\bf V}_{n,n-1}{\bf a}_{n-1}\,.
\label{eq1.19}
\end{equation} 
Here, ${\bf a}_{n}$ is the vector consisting of the $L^{d-1}$
amplitudes on the lattice sites of the cross sectional plane of the
bar at $n$, ${\bf V}_{n,n+1}$ is the $M \times M$ ($M=L^{d-1}$) dimensional
matrix of inter-layer couplings between sites on the cross sections at $n$ and
$n+1$, and ${\bf H}_{n}$ is the matrix of intra-layer couplings between sites on the cross section at $n$. 
Equation~(\ref{eq1.19}) couples
the amplitudes of a state at energy $E$ on the cross section $n+1$
to those at the cross sections $n$ and $n-1$.
We rewrite (\ref{eq1.19}) to define the $2M \times 2M$ transfer matrix,
\begin{equation}
{\bf T}_{n}=\left(
 \begin{array}{ccc} 
{\bf V}_{n,n+1}^{-1}(E{\bf 1}-{\bf H}_{n})&{,}&-{\bf V}_{n,n+1}^{-1}{\bf V}_{n,n-1}\\
{\bf 1}&{,}&{\bf 0}
\end{array}
\right) \,,
\label{eq1.20}
\end{equation}
and the transfer matrix product for the whole bar of length $N$
\begin{equation}
{\bf Q}_{N}=\prod_{n=1}^{N}\,{\bf T}_{n}\,.
\label{eq1.21}
\end{equation}  
With this, we write
\begin{equation}
\left(
\begin{array}{c} 
{\bf a}_{N+1}\\
{\bf a}_{N}
\end{array}
\right)={\bf Q}_{N}\,
\left(
 \begin{array}{c} 
{\bf a}_{1}\\
{\bf a}_{0}
\end{array}
\right)\,.
\label{eq1.22}
\end{equation}
As a consequence of Oseledec's theorem,\cite{o1968,r1982,k1985,cl1990} the eigenvalues
$\lambda_i$ of the matrix,
\begin{equation}
\Omega= \ln \left( {\bf Q}_{N} {\bf Q}_{N}^\dagger \right) \,,
\label{eq1.23}
\end{equation} 
obey the following limit
\begin{equation}
\gamma_i = \lim_{N\to\infty} \frac{\lambda_i}{2N} \,.
\label{eq1.23a}
\end{equation} 
Here, $i$ indexes the $2M$ eigenvalues of $\Omega$.
The values on the left hand side are called Lyapunov exponents. 
They occur in pairs of opposite sign.
The smallest positive Lyapunov exponent is the inverse of the quasi-one
dimensional localisation length, i.e.
\begin{equation}
\gamma_M = \frac{1}{\lambda} \,,
\label{eq1.23b}
\end{equation} 
where we have assumed that the exponents are labelled in decreasing order.

\begin{figure}
\centerline{\psfig{file=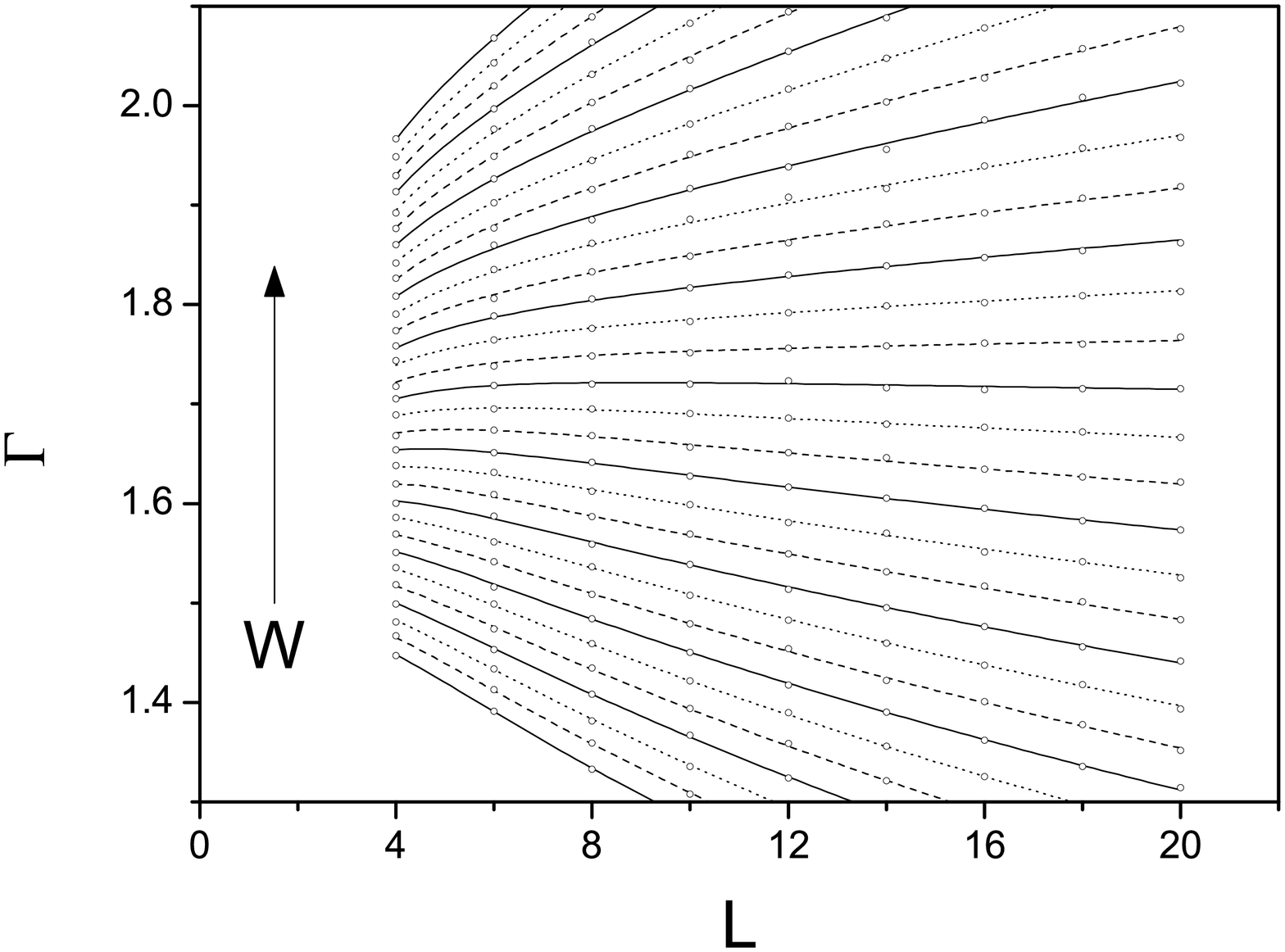,width=14cm}}
\caption{Numerical data for the three dimensional Anderson model with box distributed random potential, 
width $W=15 - 18$ in steps of $0.1$. The precision of the data is $0.1\%$. 
The lines are a finite size scaling fit that includes corrections to scaling.}
\label{fig1.1}
\end{figure}

Some typical high precision numerical data for the Anderson model in three dimensions
obtained using the transfer matrix method are shown
in Figure \ref{fig1.1}.
For weak disorder $\Gamma$ decreases, which indicates that in the three dimensional limit the system is in the
metallic phase.
For strong disorder $\Gamma$ increases, which indicates that in the three dimensional limit the system is now in the
localised phase.
At the critical disorder, we see that $\Gamma$ is independent of system size.
Note that a transient behaviour for small system sizes is clearly resolved, which must be taken into account
by including corrections to scaling when fitting the numerical data.

\subsection{The correlation length}

In addition to the critical exponent $\nu$ and the scaling functions, one of the principal results 
of the finite size scaling analysis is the correlation length $\xi$.
We find in the localised regime that 
\begin{equation}
\lim_{L\to\infty}\,\lambda(L)\,=\xi.
\label{eq1.16}
\end{equation}
Thus, provided the system is in the localised phase,
we can identify $\xi$ with the localisation length in the infinite $d$-dimensional system.
Note that it is important to distinguish the quasi-one dimensional localisation $\lambda$ on a long bar, which
is always finite,
from the localisation length $\xi$ in the the infinite $d$-dimensional system, which diverges at the Anderson
transition. 
Equation (\ref{eq1.16}) applies only in the localised phase.

Physically the localisation length $\xi$ describes the exponential decay of the transmission probability
$t(E;{\bf x},{\bf x'})$ of a quantum particle between two sites ${\bf x}$ and ${\bf x'}$ in an infinite
$d$-dimensional system that is in the localised phase
\begin{equation}
\frac{2}{\xi}=-\,\lim_{\mid {\bf x}-{\bf x'}\mid \to \infty}\,
\frac{\left< \ln t(E;{\bf x},{\bf x'}) \right>}{\mid {\bf x}-{\bf x'}\mid} \,.
\label{eq1.8}
\end{equation}
Thus, the transmission probability, and hence the diffusion constant, vanish in
the thermodynamic limit and the system is an insulator.\cite{km1993}

In the metallic phase, the correlation length is again finite and can be related to the resistivity.

\section{The critical exponents}
\label{sec1.4}

\subsection{Numerical results}
Most strikingly, although corrections to scaling had not been
considered extensively at that time, already the first works dealing
with the orthogonal symmetry class showed that the finite size scaling
method was able to confirm the most important result of the scaling theory of localisation:
whereas in three dimensions clear evidence for the existence of a critical point was found, none was
found in two dimensions.\cite{mk1981,mk1983}
During subsequent years, the universality of the critical
behaviour for the orthogonal class was explicitly demonstrated by
analysing orthogonal models with different disorder distributions.\cite{so1999}
It was also demonstrated that the high precision estimates of the critical exponents
could also be obtained by analysing the finite size scaling of various statistics
of the conductance distribution.\cite{smo2001,smo2003}

In addition, the critical behaviours of the other universality classes
have been extensively studied.
As can be seen by reference to Table 1, in a given dimension, the values of the
exponents in the different symmetry classes differ only by several percent.
Success in clearly distinguishing the critical exponents for the different universality classes 
is a triumph of the finite size scaling method.
This is in sharp contrast to other methods of estimating the exponents, in particular, the
$\epsilon$ expansion, which have singularly failed to yield precise estimates of the exponents and
even in some cases predicted values that violate the well established inequality
\cite{chayes86,kramer93a}
\begin{equation}
\nu\ge \frac{2}{d}\,.
\end{equation}

\begin{table}
\begin{center}
\begin{tabular}{|c|c|}
\hline
$\nu=1.57\pm 0.02$ & 3D orthogonal symmetry \cite{so1999}\\
\hline
$\nu=1.43\pm 0.04$ & 3D unitary symmetry \cite{so1997}\\
\hline
$\nu=1.375\pm 0.016$ & 3D symplectic symmetry \cite{asada05}\\
\hline
$\nu=2.73\pm 0.02$ & 2D symplectic symmetry \cite{aso2002}\\
\hline
$\nu=2.593\pm 0.006$ & integer quantum Hall effect \cite{slevin09}\\
\hline
\end{tabular}
\caption{List of critical exponents for different universality classes and
in different dimensions.  The error is a 95\% confidence
interval.}
\end{center}
\end{table}

\subsection{Remarks concerning experiments}

Measurement of the conductivity at finite temperature on the metallic side of the transition and 
extrapolation to zero temperature permits an estimate of $s$.
Measurement of the temperature dependence of the conductivity on the insulating side of the 
transition and fitting to the theory of variable range hopping\cite{mott,es} permits an estimate of $\nu$.

An alternative approach, called finite temperature scaling,\cite{bsb99} is to fit finite temperature 
conductivity data on both sides of the transition to
\begin{equation}
\sigma(T)=T^{s/z\nu}f(\chi /T^{1/z\nu}).
\label{eq:fts}
\end{equation}
This permits estimates of $s$ and the product $z\nu$.
Here, $\chi$ is the relevant scaling variable, which is a function of the parameter used to 
drive the transition. For example, for a transition driven by varying the carrier concentration,
we can approximate 
\begin{equation}
\chi \approx \frac{\left(n-n_\mathrm{c}\right)}{n_\mathrm{c}}\,,
\end{equation}
for doping concentrations $n$ sufficiently close to the critical concentration $n_{\mathrm c}$.
The exponent $z$, which is called the dynamical exponent, describes the divergence of the phase
coherence length as the temperature tends to zero
\begin{equation}
L_{\varphi} \sim T^{-1/z}\,.
\end{equation}
Fitting the temperature dependence of the conductivity precisely at the critical point, and
assuming the validity of Wegner's scaling law Eq.(\ref{eq:wegner}), permits an estimate of $z$.
In quantum Hall effect experiments, $z$ has been estimated by exploiting the fact that
a crossover in the temperature dependence can be observed in very small systems when the 
phase coherence length becomes comparable to the systems size.

The most recent experiments on doped semiconductors\cite{itoh04} have yielded values of $s$ and $\nu$
in the range between 1 and 1.2 that are consistent with Wegner's scaling law Eq.(\ref{eq:wegner}).
However, there is a clear deviation of the values of $\nu$ from those in Table 1.
The most recent experimental estimate of the critical exponent for the plateau transition in the
integer quantum Hall effect is $\nu=2.38\pm 0.06$.\cite{li09} 
Again this differs from the numerical estimate given in Table 1. 

The limitations of models of non-interacting electrons as a description of the critical behaviour of the
Anderson transition in electronic systems is clearly seen in the disagreement between the predicted
and measured values of the dynamical exponent $z$.
Whereas models of non-interacting electrons predict $z=d$,\cite{wegner76}
where $d$ is the dimensionality, the experimentally observed value is often smaller.
Itoh {\it et al.}\cite{itoh04} found
$z\approx 3$ in vanishing magnetic field, which agrees with non-interacting theory,
but $z\approx 2 $ in applied magnetic field, which does not.
For the plateau transition Li {\it et al.}\cite{li09} found $z\approx 1$ , which again disagrees with non-interacting theory.

The advent of experiments with cold atomic gases,\cite{chabe08}
and also with ultrasound in random elastic media,\cite{faez09}
have allowed Anderson localisation and the Anderson transition to be measured in systems that can be 
reasonably described as non-interacting.
In particular, Chabe {\it et al.}\cite{chabe08}
recently measured the critical behaviour of the Anderson transition 
in a quasi-periodic kicked rotor that was realised in a cold gas of cesium atoms.
For this system, which is 
in the three dimensional orthogonal universality class, Chabe {\it et al.} found $\nu=1.4\pm 0.3$;
a result that is
consistent with the numerical estimate in Table 1.

\section{Conclusions}
\label{sec1.5}

The finite size scaling method combined with high precision numerical simulations has
permitted the successful verification of the fundamental assumptions underlying the scaling theory
of localisation and provided high precision estimates of the critical exponents.
The advent of cold atomic gasses has permitted the experimental observation of
the Anderson transition in a system that can be reasonably described as non-interacting.
Describing the critical behaviour observed at the Anderson transition in electronic systems 
remains a challenge and would seem to require the development of numerically tractable models 
that include the long range Coulomb interaction between the electrons.

\end{document}